\documentclass[a4paper,11pt]{article}
\usepackage[a4paper, total={17cm, 25cm}]{geometry}
\usepackage{fancyhdr}
\fancypagestyle{plain}{
\fancyhf{} 
\fancyhead[RH]{\thepage}
\fancyfoot[CF]{28th Texas Symposium on Relativistic Astrophysics \\ Geneva, Switzerland -- December 13-18, 2015}

}
\begin{document}
%%%%%%%%%%%%%%%%%%%%%%%%%%%%%%%%%%%%%%%%%%%%%%%%%%%%%%%%%%%%%%%%%%%%%%%%%%%%
%%%% Title

\title{\bf Broad-band properties of flat-spectrum radio-loud narrow-line Seyfert 1 galaxies}

\author{
L. Foschini, INAF Osservatorio Astronomico di Brera, 23807 Merate, Italy 
\\
M.~Berton, Dip Fisica e Astronomia, Universit\`a di Padova, Italy
\\
A.~Caccianiga, INAF Osservatorio Astronomico di Brera, 20121 Milano, Italy
\\
S.~Ciroi, Dip Fisica e Astronomia, Universit\`a di Padova, Italy
\\
V.~Cracco, Dip Fisica e Astronomia, Universit\`a di Padova, Italy
\\
B.~M. Peterson, The Ohio State University, OH 43210, Columbus, USA
\\
E.~Angelakis, Max-Planck-Institut f\"ur Radioastronomie, 53121 Bonn, Germany
\\
V.~Braito, INAF Osservatorio Astronomico di Brera, 23807 Merate, Italy
\\
L.~Fuhrmann, Max-Planck-Institut f\"ur Radioastronomie, 53121 Bonn, Germany
\\
L.~Gallo, Dept Astronomy \& Physics, Saint Mary's University, Halifax, Canada
\\
D.~Grupe, Space Science Center, Morehead State University, Morehead, KY, USA
\\
E.~J\"arvel\"a, Aalto University, Dept Radio Science \& Engineering, Aalto, Finland
\\
S.~Kaufmann, MCTP-UNACH, Tuxtla Guti\'errez, Chiapas, M\'exico
\\
S.~Komossa, Max-Planck-Institut f\"ur Radioastronomie, 53121 Bonn, Germany
\\
Y.~Y.~Kovalev, Astro Space Center of the Lebedev Physical Institute, Moscow, Russia
\\
A.~L\"ahteenm\"aki, Aalto University, Dept Radio Science \& Engineering, Aalto, Finland
\\
M.~M.~Lisakov, Astro Space Center of the Lebedev Physical Institute, Moscow, Russia
\\
M.~L. Lister, Dept Physics \& Astronomy, Purdue Univ., West Lafayette, IN, USA
\\
S.~Mathur, The Ohio State University, OH 43210, Columbus, USA
\\
J.~L.~Richards, Dept Physics \& Astronomy, Purdue Univ., West Lafayette, IN, USA
\\
P.~Romano, INAF Istituto di Astrofisica Spaziale e Fisica Cosmica, Palermo, Italy
\\
A.~Sievers, Institut de Radio Astronomie Millim\'etrique, Granada, Spain
\\
G.~Tagliaferri, INAF Osservatorio Astronomico di Brera, 23807 Merate, Italy
\\
J.~Tammi, Aalto University Mets\"ahovi Radio Observatory, Kylm\"al\"a, Finland
\\
O.~Tibolla, MCTP-UNACH, Tuxtla Guti\'errez, Chiapas, M\'exico
\\
M.~Tornikoski, Aalto University Mets\"ahovi Radio Observatory, Kylm\"al\"a, Finland
\\
S.~Vercellone, INAF Istituto di Astrofisica Spaziale e Fisica Cosmica, Palermo, Italy
\\
G.~La~Mura, Dip Fisica e Astronomia, Universit\`a di Padova, Italy
\\
L.~Maraschi, INAF Osservatorio Astronomico di Brera, 20121 Milano, Italy
\\
P.~Rafanelli, Dip Fisica e Astronomia, Universit\`a di Padova, Italy
}

\date{\today}

\maketitle

\begin{abstract}
We report about recent updates of broad-band properties of radio-loud narrow-line Seyfert 1 galaxies.
\end{abstract}
%%%%%%%%%%%%%%%%%%%%%%%%%%%%%%%%%%%%%%%%%%%%%%%%%%%%%%%%%%%%%%%%%%%%%%%%%%%%

%%%% Paper body

\section{Introduction}
The discovery of high-energy $\gamma$ rays from radio-loud narrow-line Seyfert 1 galaxies (RLNLS1, \cite{LAT1, LAT2, LAT3, LAT4}) shook our notions  about powerful relativistic jets. The first striking implication was about jet formation and development: narrow-line Seyfert 1 galaxies (NLS1) were usually thought to have small mass central black holes, very high accretion luminosity, and negligible or even absent radio emission, although a small percent of them were known to be radio loud (e.g. \cite{BOROSON, GRUPE00, KOMOSSA}). These properties are quite at odds with those of blazars -- the known population of active galactic nuclei (AGN) with jets viewed at small angles -- but shortly after the first $\gamma$-ray detection, it was suggested that RLNLS1s could be the low-mass/high-accretion tail of the quasar distribution \cite{LAT1}. To prove this conjecture on a statistical basis, we set up a survey collecting all the available information about a large sample of known RLNLS1s.

\section{The 2015 Multiwavelength Survey}
The 2015 Multiwavelength Survey \cite{FOSCHINI15} was made both by obtaining new observations ({\it Swift}, 25 sources; {\it XMM-Newton}, 3 sources; Asiago Astrophysical Observatory, 2 sources)\footnote{A MW campaign on J$0324+3410$ with {\it INTEGRAL} and {\it Swift} was organized in 2013. Preliminary results were published in Tibolla et al. \cite{TIBOLLA} and a more detailed work is in preparation.} and by collecting all the available data at any wavelength in public archives and in literature. We focused on sources with flat radio spectra ($\alpha_{\rm r}<0.5$, $S_{\nu}\propto \nu^{-\alpha_{\rm r}}$), even though we had spectral information for half of the sources only (21/42). The remaining half of the sample had only one detection at 1.4~GHz. One of these sources, J$0953+2836$, was later detected also at 9~GHz, revealing a flat spectrum \cite{RICHARDS}. A summary of the study of steep-spectrum RLNLS1s (when the jet is likely observed at large angles, i.e. the parent population) can be found in Berton et al. \cite{BERTONTS}.

The main result of the 2015 MW Survey is to confirm that RLNLS1s are the low-mass/high-accretion tail of the quasar distribution. Although there are some observational differences in spectra, variability, MW properties, spectral energy distributions, the jets of RLNLS1s seem to be the same as blazar jets, but rescaled according to the smaller mass of the central black hole. Once properly renormalised, the jet powers of blazars, RLNLS1s, and also Galactic binaries are all consistent with each other \cite{FOSCHINI14}. The low-mass tail conjecture was also confirmed by Berton et al. \cite{BERTONLF} by means of the radio luminosity function (see Fig.~4 of \cite{BERTONLF}). 

The low luminosities of RLNLS1 made it difficult to detect this type of source at any wavelength. Future facilities -- for example, the Square Kilometre Array (SKA) -- could increase by about two orders of magnitudes the size of the RLNLS1s sample \cite{BERTONSKA}. 

Another hypothesis to explain the small number of known RLNLS1s requires an intermittent jet triggered by disk instabilities \cite{GUCHEN, DOI, FOSCHINI15}, which in turn could provide a link with compact steep spectrum (CSS) radio sources \cite{KOMOSSA, GALLO, CACCIANIGA, BERTONLF}. With reference to the intermittent jet, it is worth mentioning that X-ray studies of the radio-quiet NLS1, Mrk 335, seem to indicate there are times when the primary X-ray emission becomes collimated and is moving away from the disc.  These observations could be suggesting that the NLS1 is attempting to launch a jet base in the X-ray regime close to the black hole \cite{GALLO2, WILKINS1, WILKINS2}.

\section{Update: $\mathbf \gamma$-ray detections}
We would like to take this opportunity to report about new research published after our 2015 MW Survey \cite{FOSCHINI15} and, in particular, discuss individual sources in our list. There was one more detection at high-energy $\gamma$ rays, specifically the source J$1644+2619$ \cite{DAMMANDO}, which increases the detection rate to 19\% (8/42). More RLNLS1s -- not in our sample -- were detected at GeV energies: B3~$1441+476$ \cite{LIAO},  SDSS~J$122222.55+041315.7$ \cite{YAO}, and perhaps RX~J$2314.9+2243$ \cite{KOMOSSA15}\footnote{The latter did not perform the analysis of $\gamma$-ray data and relied on an unpublished preliminary result by Foschini.}. Detections related to the candidates for the parent population are reported by Berton et al. \cite{BERTONTS}.

The general approach to identifying RLNLS1s has been through optical spectroscopy and radio emission, and then to search for high-energy $\gamma$ rays, as was done since the early detections \cite{LAT1,LAT3,FOSCHINI11}. It seems that none followed the opposite way, i.e. to search among the unidentified or poorly known $\gamma$-ray sources of the {\it Fermi} LAT catalogs. It is not a matter to search for RLNLS1s only: it is now evident that the environment nearby the central black hole has impact on how the jet appears, but not on its formation\footnote{This is not so strange if we remind what R. Blandford said to reply to a question by G. Burbidge at the Pittsburgh Conference on BL Lac Objects in 1978 \cite{BLANDFORD78}: ``As the continuum emission is proposed to originate in the central 10 pc, I don't think the nature of the surrounding object is particularly relevant to the model''. In addition, there is the Livio's conjecture \cite{LIVIO}: ``I will make the assumption that the jet formation mechanism, namely, the mechanism for acceleration and collimation, is the same in all of the different classes of objects which exhibit jets... It should be noted right away that the emission mechanisms which render jets visible in the different classes of objects, are very different in objects like, for example, YSOs and AGN''.}. Our 2015 MW Survey favours this hypothesis. 

Instead, to understand the engine physics, particularly with reference to the unification of relativistic jets, it is necessary to observe the low-mass tail of the quasar distribution, which was neglected in the past because of the belief that powerful relativistic jets in AGN required a threshold mass of the compact object (e.g. \cite{LAOR}). Therefore, searching for small-mass quasars, we noted that the optical follow-up of the {\it Fermi} 1LAC Catalog contains a little treasure \cite{SHAW}. It reports line widths and black hole masses for 229 $\gamma$-ray detected quasars, with new spectroscopy for 165 sources. The masses were calculated by using the virial method and the FWHM of H$\beta$, Mg II, or C IV. We searched for quasars with $M<10^{8}M_{\odot}$ and the results are summarized in Table~\ref{tab:shaw}. Not every source has both measurements for two emission lines. In those cases when it was possible, we found 9 quasars at $z<1$ with two measurements both in agreement on values below $10^8M_{\odot}$ and, interestingly, 3 quasars at $z>1$. 

\begin{table}[ht]
\caption{Number of $\gamma$-ray quasars from \cite{SHAW} with $M<10^{8}M_{\odot}$ as estimated from at least one line width. The last column indicates the number of sources for which both measurements agree on a value smaller than $10^{8}M_{\odot}$.}
\begin{center}
\begin{tabular}{ccccc}
\hline
Sample & $M_{\rm H\beta}$ & $M_{\rm Mg\, II}$ & $M_{\rm C\, IV}$  & Both\\
\hline
$z<1$ (109 objects) & 22 (20\%)           & 21 (19\%)      & --       & 9 (8\%) \\
$z>1$ (119 objects) & --                  & 15 (13\%)      & 9 (7.6\%) & 3 (2.5\%) \\
\hline
\end{tabular}
\end{center}
\label{tab:shaw}
\end{table}%

The 1LAC source with the smallest mass, $3.2\times 10^{6}M_{\odot}$, is J$0430-2507$ ($z=0.516$), calculated from the Mg II line width (FWHM$= 1500\pm 600$ km~s$^{-1}$). This object is classified as low-frequency peaked BL Lac object, because of the small equivalent width of the line ($\sim0.9$~\AA \, \cite{SHAW}). Therefore, there could be doubt that the narrowness of the emission line could be due to the overwhelming jet activity (see, e.g., \cite{FOSCHINI12}). Obviously, more detailed reanalysis of the spectra is advisable. Indeed, the analysis pipeline in \cite{SHAW} was set up having in mind only quasars and BL Lacs, but not RLNLS1s. Nevertheless, there are three RLNLS1s in the 1LAC sample listed as flat-spectrum radio quasars:  

\begin{itemize}
\item J$0948+0022$: Shaw et al. \cite{SHAW} reported FWHM(H$\beta$)$=1800\pm400$~km~s$^{-1}$ and FWHM(Mg II)$=2400\pm300$~km~s$^{-1}$, which led to $M_{\rm H\beta}\sim 2.9\times 10^{7}M_{\odot}$ and $M_{\rm Mg\, II}\sim 5.4\times 10^{7}M_{\odot}$. These values can be compared with ours \cite{FOSCHINI15}: FWHM(H$\beta$)$\sim 1639$~km~s$^{-1}$ and $M_{\rm H\beta}\sim 7.5\times 10^{7}M_{\odot}$ (please note that our mass estimates were based on the line dispersion $\sigma$).

\item J$1222+0413$: Shaw et al. \cite{SHAW} found only the magnesium line, FWHM(Mg II)$=2600\pm 200$~km~s$^{-1}$, which implies $M_{\rm Mg\, II}\sim 2.3\times 10^{8}M_{\odot}$. Yao et al. \cite{YAO}, extending the optical spectrum to the rest-frame red (observed infrared), detected faint [OIII], strong Fe II, and H$\beta$ emission. The measured FWHM(H$\beta$)$=1734\pm104$ km~s$^{-1}$, which returned $M_{\rm H\beta}\sim 1.8\times 10^{8}M_{\odot}$. 

\item J$1505+0326$: the pipeline by \cite{SHAW} applied to the SDSS DR7 spectrum did not find the H$\beta$ line, although it is quite evident\footnote{{\tt http://cas.sdss.org/dr7/en/tools/explore/obj.asp?ra=226.276988\&dec=3.441892}}. The mass estimate based on magnesium line is $M_{\rm Mg\, II}\sim 2.6\times 10^{7}M_{\odot}$, in agreement with our value $M_{\rm H\beta}\sim 1.9\times 10^{7}M_{\odot}$ \cite{FOSCHINI15}. 
\end{itemize}

To conclude, even by taking into account the issues usually associated with any automatic data processing, the work by Shaw et al. \cite{SHAW} clearly indicates the presence of something to be studied in detail. Work is in progress to better assess these observations.  

\section{Update: mid-infrared observations}
A detailed study of the mid-infrared properties of the sample has been done by Caccianiga et al. \cite{CACCIANIGA15}. They analysed four-filters data (3.4, 4.6, 12, 22 $\mu$m) from the {\it Wide-field Infrared Survey Explorer} ({\it WISE}) and found that the contribution of the host galaxy is significant at 12 and 22~$\mu$m. About half of the sources displayed colours in agreement with starburst activity, which is the most important component in about 10 sources. They concluded that mid-infrared properties of the RLNLS1s are the result of at least three different contributions: jet, torus, and starburst. The variability of the jet is the key to understand what is the dominant contribution at a certain epoch.

\section{Update: radio observations}
Recently, Lister et al. \cite{LISTER} studied the pc-scale jet kinematics of a large sample of radio-loud AGN, including five RLNSL1s detected at GeV energies. Four of these sources are also in our sample and three of them displayed a highly superluminal jet motion: J$0324+3410$, $v=(9.0\pm0.3)c$; J$0849+5108$, $v=(5.8\pm0.9)c$; J$0948+0022$, $v=(11.5\pm1.5)c$ \cite{LISTER}. The fourth, J$1505+0326$, has $v=(1.1\pm0.4)c$, but it is worth noting that this source did not display significant activity since its detection (e.g. \cite{PALIYA}) and it underwent its first outburst only in December 2015 \cite{DAMCIP}.

J$0324+3410$ was also studied by Fuhrmann et al. \cite{FUHRMANN}, who reported $v=(1\div7)c$.

Another recent study on the pc-scale jet kinematics of a sample of 14 RLNLS1s was done by \cite{GU15}, who focussed on sources not detected at $\gamma$ rays and found compact core-jet structures and core brightness temperatures up to $\sim 10^{11}$~K.

It is also worth noting the recent discovery of a radio jet of Fanaroff-Riley type I in the narrow-line Seyfert 1 galaxy Mrk 1239, a (radio quiet) source characterized by a small mass of the central black hole ($\sim 10^6M_{\odot}$) and low accretion luminosity ($\sim 0.02L_{\rm Edd}$) \cite{DOI15}. This source should be compared with J$2007-4434$, the outlier of our 2015 MW Survey ($M\sim 4.3\times 10^6M_{\odot}$, $L_{\rm disk}\sim 0.003L_{\rm Edd}$, \cite{FOSCHINI15}), making it less unusual and worth studying.

%%%% References

%%%%%%%%%%%%%%%%%%%%%%%%%%%%%%%%%%%%%%%%%%%%%%%%%%%%%%%%%%%%%%%%%%%%%%%%%%%%

\begin{thebibliography}{99}

\bibitem{LAT1} Abdo, A.~A. et al. (LAT Collaboration) (2009) ApJ 699, 976
\bibitem{LAT2} Abdo, A.~A. et al. (LAT Collaboration) (2009) ApJ 707, 727
\bibitem{LAT3} Abdo, A.~A. et al. (LAT Collaboration) (2009) ApJ 707, L142
\bibitem{BERTONLF} Berton, M., Caccianiga, A., Foschini, L. et al. (2016) {\tt arXiv:1601.06165}
\bibitem{BERTONSKA} Berton, M., Foschini, L., Caccianiga, A. et al. (2016) {\tt arXiv:1601.05791}
\bibitem{BERTONTS} Berton, M. et al. (2016) These proceedings
\bibitem{BLANDFORD78} Blandford, R.~D., \& Rees, M.~J. (1978) in: Pittsburgh Conference on BL Lac Objects, University of Pittsburgh, p. 328
\bibitem{BOROSON} Boroson, T.~A. (2002) ApJ 565, 78
\bibitem{CACCIANIGA} Caccianiga, A., Ant\'on, S., Ballo, L. et al. (2014) MNRAS 441, 172
\bibitem{CACCIANIGA15} Caccianiga, A., Ant\'on, S., Ballo, L. et al. (2015) MNRAS 451, 1795
\bibitem{DAMMANDO} D'Ammando, F., et al. (2015), MNRAS, 452, 520
\bibitem{DAMCIP} D'Ammando, F., Ciprini, S., et al. (LAT Collaboration) (2015) ATel 8447
\bibitem{DOI} Doi, A., Nagira, H., Kawakatu, N., et al. (2012) ApJ 760, 41
\bibitem{DOI15} Doi, A., Wajima, K., Hagiwara, Y., \& Inoue, M. (2015) ApJ 798, L30
\bibitem{FOSCHINI11} Foschini, L. (2011): in: Narrow-Line Seyfert 1 Galaxies and Their Place in the Universe, eds L. Foschini, M. Colpi, L. Gallo et al., Proceedings of Science NLS1, id 24
\bibitem{FOSCHINI12} Foschini, L. (2012) Res Astron Astrophys 12, 359
\bibitem{FOSCHINI14} Foschini, L. (2014) IJMP CS 28, 1460188
\bibitem{LAT4} Foschini, L., et al. (LAT Collaboration) (2010) in: Accretion and Ejection in AGNs: A Global View, L. Maraschi, G. Ghisellini, R. Della Ceca, and F. Tavecchio, eds., ASP Conference Series (San Francisco, CA, USA), Vol. 427, p. 243
\bibitem{FOSCHINI15} Foschini, L., Berton, M., Caccianiga, A. et al. (2015) A\&A 575, A13
\bibitem{FUHRMANN} Fuhrmann, L., Karamanavis, V., Komossa, S. et al. (2016) MNRAS, submitted
\bibitem{GALLO} Gallo, L.~C., Edwards, P.~G., Ferrero, E., et al. (2006) MNRAS 370, 245
\bibitem{GALLO2} Gallo, L. C., Wilkins, D. R., Bonson, K. et al. (2015) MNRAS 446, 633
\bibitem{GRUPE00} Grupe, D. (2000) New Astron. Rev. 44, 455 
\bibitem{GUCHEN} Gu, M., \& Chen, Y. (2010) AJ 139, 2612
\bibitem{GU15} Gu, M., Chen, Y., Komossa, S. et al. (2015) ApJS 221, 3
\bibitem{KOMOSSA} Komossa, S., Voges, W., Xu, D., et al. (2006) AJ 132, 531
\bibitem{KOMOSSA15} Komossa, S., Xu, D., Fuhrmann, L., et al., (2015), A\&A, 574, A121
\bibitem{LAOR} Laor, A. (2000) ApJ 543, L111
\bibitem{LIAO} Liao, N.-H., Liang, Y.-F., Weng, S.-S., et al., (2015), {\tt arXiv:1510.05584}
\bibitem{LISTER} Lister, M. L., Aller, M. F., Aller, H. D. et al. (2016) AJ, submitted
\bibitem{LIVIO} Livio, M. (1997) in: Accretion Phenomena and Related Outflows, IAU Colloquium 163, eds D.~T. Wickramasinghe, L. Ferrario, \& G.~V. Bicknell, ASP Conf. Series 121, 845
\bibitem{PALIYA} Paliya, V. S., Stalin, C. S. \& Ravikumar, C. D. (2015) AJ 149, 41
\bibitem{RICHARDS} Richards, J. L., \& Lister, M. L. (2015) ApJ 800, L8
\bibitem{SHAW} Shaw, M. S., Romani, R. W., Cotter, G. et al. (2012) ApJ 748, 49
\bibitem{TIBOLLA} Tibolla, O., Kaufmann, S., Foschini, L. et al. (2013) in: Proceedings of the $33^{\rm rd}$ International Cosmic Rays Conference, 1107 ({\tt arXiv:1306.4017})
\bibitem{WILKINS2} Wilkins, D. R. \& Gallo, L. C. (2015) MNRAS 449, 129
\bibitem{WILKINS1} Wilkins, D. R., Gallo, L. C., Grupe, D. et al. (2015) MNRAS 454, 4440
\bibitem{YAO} Yao, S., Yuan, W., Zhou, H., et al., (2015), MNRAS, 454, L16
\end{thebibliography}
\end{document}